\documentclass[]{elsarticle}
\usepackage{lineno,hyperref}
\usepackage{epstopdf}
\usepackage{color}
\usepackage{lscape}
\usepackage[utf8]{inputenc}
\usepackage[T1]{fontenc}
\modulolinenumbers[5]
\biboptions{sort&compress}
\usepackage{graphicx}
\usepackage{amssymb}
\begin{document}
\begin{frontmatter}
\title{Relationship of structural disorder and stability of supercooled liquid state with glass-forming ability of metallic glasses }

\author[NWPU]{J.B. Cui}
\author[VSPU]{R.A. Konchakov}
\author[VSPU]{G.V. Afonin}
\author[VSPU]{A.S. Makarov}
\author[NWPU]{G.J. Lyu}
\author[NWPU]{J.C. Qiao}
\author[IFTT]{N.P. Kobelev}
\author[VSPU]{V.A. Khonik}
\corref{cor}
\cortext[cor]{Corresponding author} 
\ead{v.a.khonik@yandex.ru} 
\address[NWPU] {School of Mechanics, Civil Engineering and Architecture, Northwestern Polytechnical University, Xi’an 710072, China}
\address[VSPU] {Department of General Physics, Voronezh State Pedagogical
University,  Lenin St. 86, Voronezh 394043, Russia}

\address[IFTT] {Institute of Solid State Physics RAS, Chernogolovka, 142432, Russia}

\begin{abstract}
We performed calorimetric studies of 26 metallic glasses and calculated the excess entropy and excess enthalpy  with respect to their counterpart crystals. On this basis, we introduced a dimensionless entropy-based parameter $\sigma_{scl}$, which characterizes structural disordering and stability of the supercooled liquid state upon heating. A very good correlation of $\sigma_{scl}$ with literature data on the critical cooling rate $R_c$ and critical diameter $D_{max}$ of metallic glasses is shown. We also introduced another dimensionless parameter  $\eta_{scl}$ based on the excess enthalpy of glass and showed that $\eta_{scl}$ provides equally good correlation with $R_c$ and $D_{max}$. Possible relationship  of structural disordering and glass-forming ability in the supercooled liquid range  with the defect structure of glass is discussed. The obtained results provide a new window for the understanding of the glass-forming ability of metallic glasses.  
\end{abstract}


\end{frontmatter}

\section{Introduction}  Melts of different materials can be cooled below the equilibrium solidus point. If the undercooling is  significant, the melt freezes and forms a glass.  The prediction and physical interpretation of the glass-forming ability (GFA) of undercooled melts is one of major problems in the   physics of glassy materials.  Quantitatively, the GFA is estimated by the critical cooling rate $R_c$, which is the minimal rate of melt cooling required to produce a glass. Another GFA measure is the maximal diameter of fully non-crystalline casting produced by melt cooling, which is called the critical diameter $D_{max}$. Both GFA parameters, $R_c$ and $D_{max}$, are not easy to be determined and, therefore,  quite a few indirect GFA measures have been proposed, which make use  of  a combination of characteristic temperatures, e.g. the glass transition, crystallization and liquidus temperatures \cite{ChattopadhyaMaterSciTechnol2015}.

The GFA of metallic glasses (MGs)  has been analyzed in numerous works, e.g. see the reviews \cite{ChattopadhyaMaterSciTechnol2015,LuIntermet2007, MakarovJRTPLett2024}. It is known that the GFA of metallic systems strongly depends on chemical composition so that even various minor dopants can change it strongly \cite{WangWHPMS2007}. To interpret the GFA, different approaches are used, in particular accounting for the  difference between the atomic diameters of the alloy components \cite{InoueActaMater2000}, mismatch enthalpy and entropy \cite{TakeuchiMaterTransJim2000}, melting and mixing enthalpy \cite{CaiMSE2007}, mismatch entropy and configurational entropy \cite{RaoIntermetallics2013}, etc.  \cite{ChattopadhyaMaterSciTechnol2015}. However, the thermodynamic parameters
used for GFA estimates \cite{TakeuchiMaterTransJim2000,CaiMSE2007,RaoIntermetallics2013} are the same  for both crystalline and glassy states that strongly restricts their significance.  

The entropy approach to the understanding of different MGs' properties is of a special interest
\cite{JohariJPC2019,YangJPCL2022,FengFundRes2022,LuJPCL2023}. In particular, entropy notions were recently applied to GFA analysis of MGs \cite{MakarovJRTPLett2024,MakarovScrMat2024}. First, it was shown that the GFA estimated via a number of parameters based on characteristic temperatures systematically increases with the excess entropy $\Delta S$ of glass with respect to the maternal crystalline state provided that  this entropy is calculated for the supercooled liquid (SCL) state \cite{MakarovJRTPLett2024}.  Therefore, the GFA of undercooled melts increases with structural disorder in the SCL state occurring upon heating. Similar conclusion was derived by  the authors \cite{MakarovScrMat2024}, who introduced a dimensionless parameter of structural order,  $\xi(T)=1-\frac{\Delta S(T)}{S_m}$, where  $S_m$ is the melting entropy. A study performed on several MGs showed  that  $\xi$ calculated again for the SCL state increases strongly with $R_c$ and, therefore, the GFA increases  with structural disorder controlled by the entropies $\Delta S$ and $S_m$. However, the use of  $\xi$-parameter is strongly limited because the liquidus temperatures of MGs  usually exceed 900 K and, therefore, it is impossible to determine the melting entropy $S_{m}$ using conventional calorimeters.

In the view of the above, the aim of this work is to systematically test the hypothesis that
the GFA of metallic melts is related to the excess entropy and excess enthalpy of corresponding glasses. We suggested new entropy- and enthalpy-based parameters and using the data on 26 MGs in bulk and ribbon forms showed a very good correlation of these parameters with literature data on the GFA expressed by $R_c$ and $D_{max}$.  

\section{Experimental}
\begin{table}[h]
\begin{center}
\footnotesize
\caption{ Critical cooling rate $R_c$, critical diameter $D_{max}$, excess entropy $\Delta S$ and excess enthalpy $\Delta H$ of bulk and ribbon metallic glasses at the glass transition temperature $T_g$ and in the end of the supercooled liquid range. } 
\begin{tabular}{p{6pt}p{120pt}p{32pt}p{32pt}p{30pt}p{30pt}p{22pt}p{22pt}}
\hline
N & Glass & $R_c$ & $D_{max}$ & ${\Delta S}_{T_g}$ & ${\Delta S}_{scl}$ & ${\Delta H}_{T_g}$ & ${\Delta H}_{scl}$\\
   &       & [K/s] & [mm] & [$\frac{J}{K \times mol}$] & [$\frac{J}{K \times mol}$] & [$\frac{kJ}{mol}$] & [$\frac{kJ}{mol}$]\\
\hline
1   & Zr$_{58.5}$Nb$_{2.8}$Cu$_{15.6}$Ni$_{12.8}$Al$_{10.3}$ (ribbon)& 1.75 \cite{HaysAPL2001} & --- & 3.9 & 5.1 & 2.9 & 3.8\\
2   & Zr$_{66}$Al$_9$Cu$_{16}$Ni$_9$ (ribbon)					     & 4.10 \cite{LuIntermet2007} & ---	& 7.3 & 7.7 & 5.3 & 5.6\\
3   & Zr$_{66}$Al$_8$Cu$_{12}$Ni$_{14}$ (ribbon)					 & 9.80 \cite{LuIntermet2007} & --- & 6.5 & 7.7 & 4.7 & 5.5\\
4   & Zr$_{55}$Al$_{19}$Co$_{19}$Cu$_7$ (ribbon)					 & 16.0 \cite{MukherjeePRB2004} & --- & 3.3 & 3.7 & 2.6 & 2.9\\
5   & Zr$_{55}$Al$_{22.5}$Co$_{22.5}$ (ribbon)					 & 18.0 \cite{MukherjeePRL2005} & --- & 7.1 & 7.3 & 5.8 & 6.0\\
6   & Zr$_{66}$Al$_{8}$Cu$_7$Ni$_{19}$ (ribbon)					 & 22.7 \cite{LuIntermet2007} & ---	& 7.5 & 8.3 & 5.4 & 5.9\\
7   & Zr$_{60}$Cu$_{20}$Ni$_8$Al$_7$Hf$_3$Ti$_2$ (ribbon)				 & 33.5 \cite{XingJALCOM2004} & ---	& 5.2 & 5.7 & 3.7 & 4.1\\
8   & Zr$_{66}$Al$_8$Ni$_{26}$ (ribbon)							 & 66.6 \cite{LuIntermet2007} & ---	& 8.0 & 9.5& 5.9 & 6.1\\
9   & Zr$_{65}$Al$_{10}$Ni$_{10}$Cu$_{15}$ (bulk)					 & 4.10 \cite{WangJAP2006} & --- & 9.0  & 10.4& 6.6 & 7.5\\
10  & Zr$_{55}$Co$_{25}$Al$_{20}$ (bulk) 						 & 16.5 \cite{MukherjeePRB2004} & 2.5 \cite{LongJALCOM2009} & 6.0  & 6.4 & 4.9 & 5.2\\
11  & Zr$_{57}$Nb$_5$Al$_{10}$Cu$_{15.4}$Ni$_{12.6}$ (bulk)			 & 10.0 \cite{RyuAPLMater2017} & 11.2 \cite{ChattopadhyayMST2016} & 5.3 & 6.6 & 5.0 & 6.0\\
12a & Zr$_{52.5}$Ti$_5$Cu$_{17.9}$Ni$_{14.6}$Al$_{10}$ (bulk)		 & 4.50 \cite{XingJALCOM2004} & 10 \cite{ChattopadhyayMST2016} & 3.8 & 4.4 & 2.8 & 3.2\\
12b & Zr$_{52.5}$Ti$_5$Cu$_{17.9}$Ni$_{14.6}$Al$_{10}$ (bulk)		 & 25.0 \cite{RyuAPLMater2017} & 10 \cite{ChattopadhyayMST2016} & 3.8 & 4.4 & 2.8 & 3.2\\
13  & Pd$_{40}$Cu$_{30}$Ni$_{10}$P$_{20}$ (bulk)					 & 0.10 \cite{InoueMT1997} & 72 \cite{ChattopadhyayMST2016} & 4.4 & 6.8 & 3.0  & 4.4\\
14  & Pd$_{40}$Cu$_{30}$Ni$_{10}$P$_{20}$ (ribbon)				 & 0.10 \cite{InoueMT1997} & 72 \cite{ChattopadhyayMST2016}& 3.1 & 4.6 & 2.0 & 3.0\\
15  & Ti$_{34}$Zr$_{11}$Cu$_{47}$Ni$_8$ (ribbon)					 & 100 \cite{LuIntermet2007} & 3.0 \cite{ChattopadhyayMST2016} & 8.9 & 9.2 & 7.0 & 7.3\\
16  & Zr$_{57}$Ti$_5$Al$_{10}$Cu$_{20}$Ni$_8$ (ribbon)				 & 10.0 \cite{LuIntermet2007} & 10 \cite{ChattopadhyayMST2016}	& 4.5 & 4.9 & 3.3 & 3.6\\
17  & Zr$_{65}$Al$_{7.5}$Cu$_{17.5}$Ni$_{10}$ (ribbon) 			 & 1.50 \cite{LuIntermet2007} & 16 \cite{ChattopadhyayMST2016}	& 6.2 & 9.6 & 4.5 & 5.6\\
18  & Pd$_{40}$Ni$_{40}$P$_{20}$ (ribbon) 						 & 1.6 \cite{SenkovPRB2007} & 25 \cite{YCHuPRM2019} & 6.3 & 8.4 & 4.2 & 5.5\\
19a & Pd$_{40}$Ni$_{40}$P$_{20}$ (bulk) 						 & 1.6 \cite{SenkovPRB2007} & 25 \cite{YCHuPRM2019} & 7.5 & 9.8 & 5.1 & 6.4\\
19b & Pd$_{40}$Ni$_{40}$P$_{20}$ (bulk) 						 & 0.17 \cite{WangJAP2006} & 25 \cite{YCHuPRM2019} & 7.5 & 9.8 & 5.1 & 6.4\\
20  & Zr$_{46}$Cu$_{46}$Al$_8$ (bulk)							 & 4.0 \cite{LanAPL2014} & --- & 3.9 & 5.3 & 3.0& 4.0\\
21  & Zr$_{50}$Cu$_{50}$ (ribbon)							     & ---	 &	2.0 \cite{LongJALCOM2009} & 7.3 & 7.9& 5.3 & 5.7\\
22  & Zr$_{48}$Cu$_{34}$Ag$_8$Al$_8$Pd$_2$ (bulk)				     & ---	 & 30 \cite{DengJNCS2020} & 3.3 & 4.4& 2.5 & 3.3\\
23  & Cu$_{49}$Hf$_{42}$Al$_9$ (bulk)				             & ---	 & 10 \cite{YCHuPRM2019} & 5.7 & 6.7 & 4.8 & 5.6\\
24  & Zr$_{50}$Cu$_{40}$Al$_{10}$ (bulk)				             & ---	 & 22 \cite{InoueAM2011}  & 4.0 & 5.0 & 3.1 & 3.8\\
25  & La$_{55}$Al$_{25}$Co$_{20}$ (bulk)				             & ---	 & 3 \cite{GhorbaniSciRep2022}  & 8.5 & 9.8 & 4.4 & 5.1\\
26  & La$_{55}$Al$_{35}$Ni$_{10}$ (bulk)				             & ---	 & 5 \cite{GhorbaniSciRep2022}  & 7.8 & 8.8 & 4.4 & 4.9\\
\hline
\end{tabular}
\end{center}
\label{onecolumntable}
\end{table}

Table 1 gives a list of MGs (at.\%) under investigation, which were produced as either 2 mm thick plates  by melt suction/jet quenching into a copper mold (labeled as bulk) or as 20-40 $\mu m$ thick ribbons produced by melt spinning (labeled as ribbon). The samples were X-ray verified to be fully non-crystalline.  

Differential scanning calorimetry (DSC)  were performed using a Hitachi DSC 7020 instrument in 99.999 pure N$_2$ atmosphere on 60-70 mg samples. DSC measurements were performed according to the following protocol. The first heating was performed with an initial sample and an empty reference cell up to the temperature of complete crystallization $T_{cr}$ and subsequent cooling to room temperature. This sample was next moved to the reference DSC cell. Next, a new sample of nearly the same mass was heated up to $T_{cr}$. This procedure allows obtaining the differential heat flow $\Delta W=W_{gl}-W_{cr}$ ($W_{gl}$ and $W_{cr}$ are the heat flows coming from the glass and counterpart crystal, respectively), which was further used for data analysis. The heating rate $\dot{T}$ was 3 K/min in all cases. 

\begin{figure}[t]
\begin{center}
\includegraphics[scale=0.8]{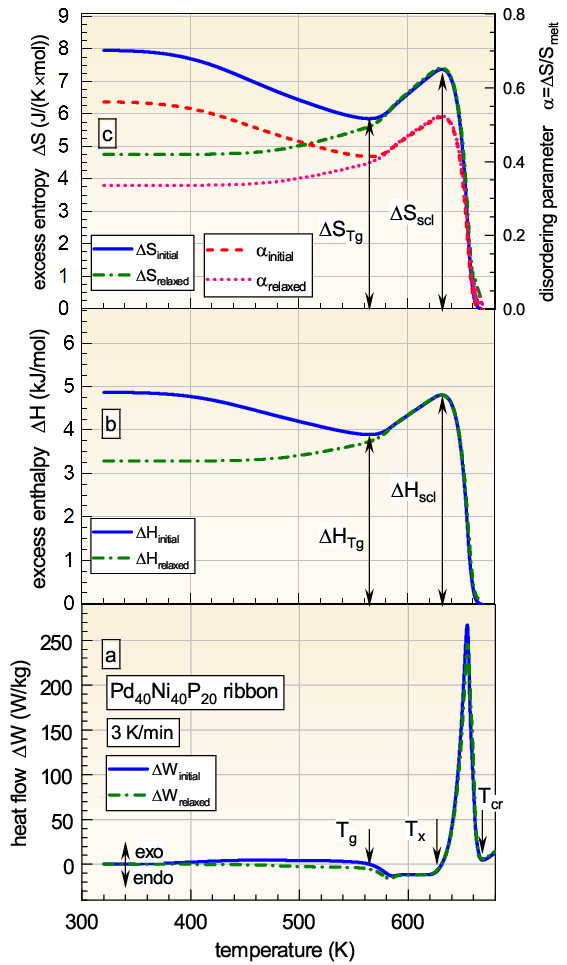}
\caption[*]{\label{Fig1.eps} Differential heat flow $\Delta W$ (a), excess enthalpy $\Delta H$ (calculated with Eq.(\ref{DeltaH})) (b), excess entropy $\Delta S$  (calculated with Eq.(\ref{DeltaS})) and disordering parameter $\alpha =\Delta S/S_{m}$ (c) for glassy ribbon  Pd$_{40}$Ni$_{40}$P$_{20}$ in the initial and relaxed states as functions of temperature. The characteristic temperatures -- glass transition $T_g$, crystallization onset $T_x$ and complete crystallization $T_{cr}$, are indicated.  The enthalpy $\Delta H_{Tg}$ and entropy $\Delta S_{Tg}$ at $T=T_g$ together with the enthalpy $\Delta H_{scl}$ and entropy $\Delta S_{scl}$ corresponding to the end of the SCL range are also shown. } 
\end{center}
\end{figure} 

The excess enthalpy $\Delta H$ and excess entropy $\Delta S$  were calculated in the way suggested recently \cite{MakarovJPCM2021,MakarovJLETT2022}:

\begin{equation}
\Delta H(T)=\frac{1}{\dot{T}}\int_T^{T_{cr}}\Delta W(T)\;dT, \label{DeltaH}
\end{equation}

\begin{equation}
\Delta S(T)=\frac{1}{\dot{T}}\int_T^{T_{cr}}\frac{\Delta W(T)}{T}\;dT. \label{DeltaS}
\end{equation}
It is seen that as temperature $T$ approaches $T_{cr}$, the integrals (\ref{DeltaH}) and (\ref{DeltaS}) tend to zero and, therefore, $\Delta H$ and $\Delta S$ represent the excess enthalpy and entropy with respect to the crystalline state.

The relaxed state of Pd$_{40}$Ni$_{40}$P$_{20}$ glass was obtained by heating into the SCL range and subsequent cooling back at 3 K/min to room temperature.

\section{Results}

Figure \ref{Fig1.eps}(a)  gives, as an example, the differential heat flow $\Delta W(T)$ for glassy ribbon Pd$_{40}$Ni$_{40}$P$_{20}$ in the initial and relaxed states, which is well-known as a good glass-former. In the initial state, small exothermal effect below the glass transition temperature $T_g$ (which is defined as the onset of the  endothermal reaction, indicated by an arrow) is determined by structural relaxation occurring upon heating. Higher temperatures belonging  to the interval $T_g<T<T_x$ ($T_x$ is crystallization onset temperature)  correspond to the SCL range, which is followed by a strong  exothermal crystallization ending at a temperature $T_{cr}$. Preliminary relaxation leads to the disappearance of the exothermal relaxation effect, as one would expect.

Panel (b) in Fig.\ref{Fig1.eps} shows  temperature dependence of the excess enthalpy  $\Delta H$ calculated using Eq.(\ref{DeltaH}) with the differential heat flow $\Delta W(T)$ shown in panel (a). In the initial state, a decrease of $\Delta H$ from 4.9 kJ/mol to 3.9 kJ/mol upon heating up to $T_g$  reflects the aforementioned exothermal structural relaxation. Entering the SCL range above $T_g$ results in rather rapid increase of the excess enthalpy up to 4.8 kJ/mol, which is related to the endothermal heat flow in the range $T_g<T<T_x$ (panel (a)). Finally, the crystallization  starting at $T_x$ results in a rapid decrease of $\Delta H$ to zero at $T_{cr}$. Relaxed sample demonstrates a significantly reduced excess enthalpy below $T_g$ but temperature dependences $\Delta H(T)$ in the initial and relaxed states at temperatures $T>T_g$ are nearly identical. Thus, heating into the SCL range removes the memory of the thermal prehistory.

All aforementioned peculiarities are further reflected in temperature evolution of the  excess entropy $\Delta S$ given in Fig.\ref{Fig1.eps}(c). Heating of an initial sample from room temperature to $T_g$ results in a significant decrease of $\Delta S$ from 8.0 J/(mol$\times $K) to 6.0 J/(mol$\times $K), which demonstrates a significant structural ordering occurring upon exothermal relaxation. In the SCL region ($T_g<T<T_x$), the excess entropy rapidly increases up to $\Delta S_{scl}$=7.4 J/(mol$\times $K) just below $T_x$ indicating a substantial disordering of the structure. The crytallization onset at $T_x$ leads to a rapid decrease in  $\Delta S$ to zero due to the complete crystallization at $T_{cr}$. Preliminary relaxation leads to a significant decrease of the excess entropy below $T_g$ but $\Delta S(T)$ dependneces at higher temperatures in the SCL range are very close.  

\begin{figure}[t]
\begin{center}
\includegraphics[scale=0.8]{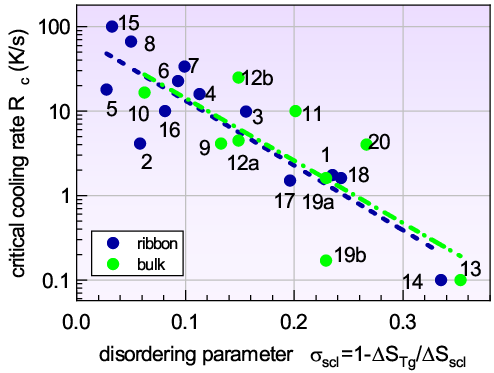}
\caption[*]{\label{Fig2.eps} Critical cooling rate $R_c$ as a function of the disordering parameter $\sigma_{scl}$ for ribbon and bulk glasses. The numbers indicate MGs' compositions according to Table 1. The lines give least square fits. The Pearson correlation coefficient is 0.86 and 0.89 for bulk and ribbon glasses, respectively.} 
\end{center}
\end{figure} 

It is further reasonable to characterize structural evolution by the disordering parameter $\alpha =\Delta S/S_{m}$  ($S_m$ is the melting entropy) \cite{AfoninJALCOM2024b}, which  changes in the range $0<\alpha<1$. The value $\alpha \rightarrow 1$ defines a liquid-like order (close to that of a liquid) and $\alpha \rightarrow 0$ correspond to a crystal-like order. Temperature dependence of $\alpha$ calculated with $S_{m}=14.2$ J/(mol$\times $K) \cite{MakarovScrMat2024} in the initial and relaxed states are given in Fig.\ref{Fig1.eps}(c). It is seen that, naturally, the character of temperature dependence of $\alpha$ repeats that of the excess entropy $\Delta S$. Preliminary relaxation provides a substantial structural ordering below $T_g$. However, it should be emphasized a significant disordering described by an increase of $\alpha$ from 0.41 to 0.52 (that is by 27\%)  upon heating in the SCL  range  and this disordering is the same for initial and relaxed states.   Nonetheless, the use of the parameter $\alpha$ is limited because of usually high liquidus temperatures, as mentioned above.  

In this work, we are most interested in the SCL range $T_g<T<T_x$. The results described above are typical for MGs and generally indicate that this range is characterized by \textit{i}) endothermal heat flow and corresponding rise of the excess enthalpy $\Delta H$ and \textit{ii}) significant structural disordering reflected by an increase of the excess entropy $\Delta S$. We show that the GFA of a \textit{melt} can be predicted using the data on the excess entropy and excess enthalpy of the \textit{glassy} state.  

\section{Discussion}

We  suggest that the GFA of a melt is related to the rise of  the excess entropy $\Delta S$ upon heating of glass in  the SCL range. This rise can be characterized by a dimensionless parameter $\sigma_{scl}=1-\frac{\Delta S_{Tg}}{\Delta S_{scl}}$, where the entropies $\Delta S_{Tg}$ and $\Delta S_{scl}$ are  defined above (see Fig.\ref{Fig1.eps}) and $\sigma_{scl}$ indicates the strength of the disordering in the SCL range.  The larger  $\sigma_{scl}$, the more disorder is introduced. On the other hand, $\sigma_{scl}$ characterizes the stability of supercooled liquid, i.e.  the larger the $\sigma_{scl}$ value, the more stable the supercooled liquid is before it crystallizes.  Therefore, one can naturally consider that more stable supercooled liquid will lead to larger undercooling upon melt quenching and, respectively, to larger  $R_c$.  This idea is verified in the present work. 

\begin{figure}[t]
\begin{center}
\includegraphics[scale=0.8]{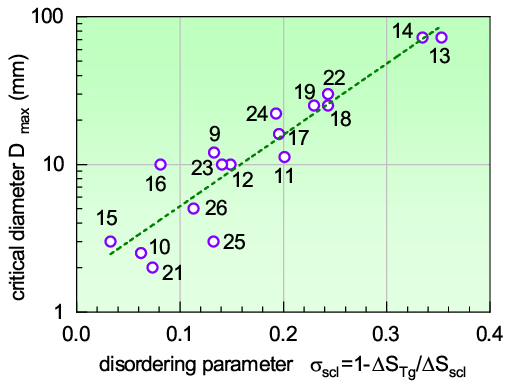}
\caption[*]{\label{Fig3.eps} Critical diameter $D_{max}$ as a function of the disordering parameter $\sigma_{scl}$ for bulk glasses.  The line gives the least square fit with a Pearson correlation coefficient of 0.94.} 
\end{center}
\end{figure} 

The entropies  $\Delta S_{Tg}$ and  $\Delta S_{scl}$  necessary for $\sigma_{scl}$-calculation together with literature data on $R_c$ and $D_{max}$  are listed in Table 1. Figure \ref{Fig2.eps} shows the critical cooling rate $R_c$ as a function of the disordering parameter $\sigma_{scl}$ for MGs in bulk and ribbon forms. It is seen that log$R_c$ steadily decreases (i.e. the GFA increases) as $\sigma_{scl}$ increases, and the data scatter is quite acceptable. The best GFA among the glasses listed in Table 1 corresponds to Pd$_{40}$Cu$_{30}$Ni$_{10}$P$_{20}$  (lines 13,14) with $R_c\approx 0.1$ K/s. Respectively,  this glass has the largest $\sigma_{scl}$ of about 0.35. The next best GFA is demonstrated by Pd$_{40}$Ni$_{40}$P$_{20}$  (lines 18,19a,19b) and Zr$_{46}$Cu$_{46}$Al$_8$. These glasses have $\sigma_{scl}$, which is a little smaller than that for Pd$_{40}$Cu$_{30}$Ni$_{10}$P$_{20}$. 

It is also worthy of note that the data for bulk and ribbon glasses do not vary much and the corresponding linear fits are quite close. Bulk and ribbon samples differ by orders of magnitude in  the melt quenching rate upon glass production that leads to differences in their structures. However, it is known (and verified above in Fig.1) that MGs in the SCL range lose the memory of the thermal prehistory because of small relaxation time, which rapidly decreases with temperature. It is for this reason that  $R_c(\sigma_{scl})$-dependences for bulk and ribbon MGs are close to each other.   

Figure \ref{Fig3.eps} gives the critical diameter $D_{max}$ as a function of $\sigma_{scl}$. Larger $D_{max}$  corresponds to larger GFA. It is seen that log$D_{max}$ increases with $\sigma_{scl}$. Again, the best glass-former, Pd$_{40}$Cu$_{30}$Ni$_{10}$P$_{20}$, which has the largest $D_{max}=72$ mm, also has the largest $\sigma_{scl}\approx 0.35$. On the other hand, the glasses having $D_{max}$ in the range of 1-2 mm, demonstrate the smallest $\sigma_{scl}$ of about 0.03.    

Thus, the data in Fig.\ref{Fig2.eps} and Fig.\ref{Fig3.eps} convincingly show that the critical cooling rate $R_c$ and critical diameter $D_{max}$, which constitute direct measures of the  GFA, display unambiguous dependence on structural disordering and melt stability in the SCL range given by the dimensionless parameter $\sigma_{scl}$. This conclusion  agrees with a fundamental work by Li et al. \cite{LiNatureMater2022} who fabricated over five thousand alloys and found a strong correlation between the high GFA and a direct measure of glass disorder given by the width (FWHM) of X-ray diffraction structure factor. On the other hand, the FWHM increases with temperature within the SCL range \cite{NeuberActamater2021} providing thus  direct confirmation of structural disordering in this range.

\begin{figure}[t]
\begin{center}
\includegraphics[scale=0.8]{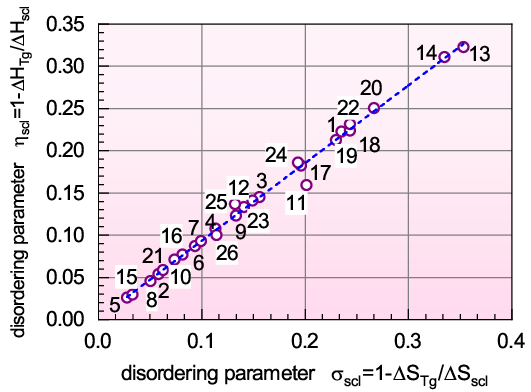}
\caption[*]{\label{Fig4.eps} Relationship between the  enthalpy-based $\eta_{scl}$ and  entropy-based $\sigma_{scl}$  parameters of structural disorder and stability in the supercooled liquid range. The slope $d\eta_{scl}/d\sigma_{scl}=0.92\pm 0.01$.}
\end{center}
\end{figure} 

Structural disordering in the SCL region should also lead to an increase in the heat content, or the enthalpy, due to the second law of thermodynamics, $dH=TdS$ for constant pressure. One can expect, therefore, that an increase of the excess enthalpy should be proportional to the increase of the excess entropy. In this case,  the rise of structural disorder and melt stability in the SCL region can be alternatively described using an enthalpy-based dimensionless parameter $\eta_{scl}=1-\frac{\Delta H_{Tg}}{\Delta H_{scl}}$, where the excess enthalpies $\Delta H_{Tg}$ and $\Delta H_{scl}$ correspond to the glass transition temperature $T_g$ and the end of the SCL range (see Fig.\ref{Fig1.eps}). These enthalpies are listed in Table 1. Figure \ref{Fig4.eps} gives $\eta_{scl}$ as a function of $\sigma_{scl}$.  A very good linear relationship between these parameters is observed. The derivative $d\eta_{scl}/d\sigma_{scl}$ in this plot is $0.92\pm 0.01$. This means that $\eta_{scl}$ is quite close to $\sigma_{scl}$ and that is why it can also be used to describe the structural disordering and stability of a supercooled melt. Thus, while the parameters $\sigma_{scl}$ and $\eta_{scl}$ are based on qualitatively different characteristics of the SCL state, namely its structural disorder and heat content, they provide equally meaningful information about this state. 

It is also important to emphasize that the $\sigma_{scl}$ and $\eta_{scl}$  can be determined using conventional calorimeters, as one does not need to reach usually high liquidus temperatures. It should also be pointed out that just a couple routine DSC measurements are necessary to estimate the GFA of a metallic glass.
 
One can suggest a qualitative interpretation of the relationship between the GFA and structural disordering in the SCL range. According to the Interstitialcy theory (IT), melting is related to   a rapid multiplication of interstitial defects in the dumbbell form (called interstitialcies). These defects remain identifiable structural units in the liquid state and melt quenching  freezes a part of them in solid  glass \cite{KobelevUFN2023,GranatoEurJPhys2014}. Then, relaxation phenomena in MGs are determined by the changes in the defect concentration. This general approach provides quantitative interpretation of  numerous relaxation effects in MGs, as described in a recent review \cite{KobelevUFN2023}. 

Meanwhile, it was shown that it is the interstitial-type defect system, which almost completely determines  the excess entropy $\Delta S$ and excess enthalpy $\Delta H$ of MGs with respect to their counterpart crystals \cite{MakarovJLETT2022,KobelevUFN2023}. Indeed,  $\Delta H$  simply constitutes the elastic energy of the defect system accurate to a precision of 10--15\% \cite{MakarovJLETT2022}. On the other hand, the excess enthalpy $\Delta H$ is proportional to the excess entropy $\Delta S$ of glass \cite{AfoninAPL2024} (this also follows from Fig.\ref{Fig4.eps}) and, therefore, the latter is determined by glass defect structure as well. 

Interstitial-type defects increase the vibrational entropy of glass through their strong low-frequency vibrational modes \cite{KobelevUFN2023,GranatoEurJPhys2014}. Structural disordering due to increasing defect concentration  leads to a rise of the vibrational component of the entropy. Thus, the total excess entropy of glass $\Delta S$ increases with the defect concentration while the latter increases with temperature in the SCL range \cite{KobelevUFN2023,MakarovIntermetallics2023}. This should result in an increase of $\Delta S$ and  $\Delta H$ in the SCL range, as indeed observed (see Fig.\ref{Fig1.eps}).

On the other hand, an increase of the defect concentration in the SCL range leads  to simultaneous increase of structural disorder given by the width (FWHM) of X-ray structure factor \cite{MakarovIntermetallics2023}. Since the increase of the normalized FWHM  with temperature approximately equals to the rise of the defect concentration \cite{MakarovIntermetallics2023}, one should accept that structural disordering in the SCL range quantified by an increase of the FWHM and $\sigma_{scl}$ is mainly caused by an increase in the defect concentration. Consequently, given that  $\sigma_{scl}$ also reflects the structural stability in the SCL range, one can expect that the GFA of \textit{undercooled melt} is governed by the defect concentration in the SCL state of \textit{glass} as well.  This is an intriguing direction for further research since it provides a completely new window to the understanding of the GFA of metallic glasses.

\section{Conclusions} 
We performed differential scanning calorimetry on 26 metallic glasses in bulk and ribbon forms. On this basis, temperature dependences of  the excess entropy and excess enthalpy of these glasses with respect to their counterpart crystals are calculated. Using  the excess entropy, we introduced a dimensionless parameter $\sigma_{scl}$, which characterizes the structural disordering and structural stability of the supercooled liquid (SCL) state. It is found that $\sigma_{scl}$ strongly correlates with the critical cooling rate $R_c$ and critical diameter $D_{max}$, which can be achieved upon glass production. Alternatively, we introduced a dimensionless parameter $\eta_{scl}$ based on the excess enthalpy in the supercooled liquid state, which equally good correlates with $R_c$ and  $D_{max}$. 

Our investigation provides convincing arguments that glass-forming ability (GFA) of a melt characterized by $R_c$ and  $D_{max}$  is intrinsically related to the degree of structural disorder, structural stability and heat content achieved upon heating above the glass transition temperature before the crystallization takes place. Possible relationship of these findings with the defect structure of metallic glasses is discussed.  It is argued that the GFA of an undercooled melt is related to the defect concentration in the SCL state of glass.

The parameters $\sigma_{scl}$ and $\eta_{scl}$ introduced in this work are very convenient for an express estimate of the GFA.

\section{Author Contributions}

J.B.C. and G.J.L. contributed to the preparation of samples, R.A.K., A.S.M. and G.V.A. performed DSC measurements and data analysis, J.C.Q., N.P.K. and V.A.K. prepared and edited the manuscript, V.A.K. conceived the idea and supervised the project. 

\section{Acknowledgments}

The work was supported by Russian Science Foundation under the project 23-12-00162.

\end{document}